\documentclass[12pt,a4paper,twoside,english]{article}
\usepackage[T1]{fontenc}
\usepackage[latin9]{inputenc}
\usepackage{color}
\usepackage{babel}
\usepackage{units}
\usepackage{amssymb}
\usepackage[unicode=true,
 bookmarks=true,bookmarksnumbered=true,bookmarksopen=true,bookmarksopenlevel=1,
 breaklinks=false,pdfborder={0 0 1},backref=false,colorlinks=true]
 {hyperref}
\hypersetup{pdftitle={The LyX Tutorial},
 pdfauthor={LyX Team},
 pdfsubject={LyX-documentation Tutorial},
 pdfkeywords={LyX, documentation},
 linkcolor=black, citecolor=black, urlcolor=blue, filecolor=blue,pdfpagelayout=OneColumn, pdfnewwindow=true, pdfstartview=XYZ, plainpages=false}
\usepackage{breakurl}

\makeatletter

\usepackage{ifpdf} 
\ifpdf 

 \IfFileExists{lmodern.sty}{\usepackage{lmodern}}{}

\fi 

\let\myTOC\tableofcontents
\renewcommand\tableofcontents{%
  \frontmatter
  \pdfbookmark[1]{\contentsname}{}
  \myTOC
  \mainmatter }

\def\LyX{\texorpdfstring{%
  L\kern-.1667em\lower.25em\hbox{Y}\kern-.125emX\@}
  {LyX}}

\makeatother

\begin{document}

\title{Spinor Geometry}

\author{A. Nicolaidis and V. Kiosses\medskip \\{\normalsize Theoretical
Physics Department}\\{\normalsize University of Thessaloniki}\\{\normalsize 54124
Thessaloniki, Greece}\\{\normalsize nicolaid@auth.gr}}
\maketitle
\begin{abstract}
It has been proposed that quantum mechanics and string theory share
a common inner syntax, the relational logic of C. S. Peirce. Along
this line of thought we consider the relations represented by spinors.
Spinor composition leads to the emergence of Minkowski spacetime.
Inversely the Minkowski spacetime is istantiated by the Weyl spinors,
while the merge of two Weyl spinors gives rise to a Dirac spinor.
Our analysis is applied also to the string geometry. The string constraints
are represented by real spinors, which create a parametrization of
the string worldsheet identical to the Enneper-Weierstass representation
of minimal surfaces. Further, a spinorial study of the $AdS_{3}$
spacetime reveals a Hopf fibration $AdS_{3}\rightarrow AdS_{2}$.
The conformal symmetry inherent in $AdS_{3}$ is pointed out. Our
work indicates the hidden ties between logic-quantum mechanics-string
theory-geometry and vindicates the Wheeler's proposal of pregeometry
as a large network of logical propositions.
\end{abstract}

\subsubsection*{Introduction}

\qquad{}Discerning the foundations of a theory is not simply a curiosity.
It is a quest for the internal architecture of the theory, offering
a better comprehension of the entire theoretical construction and
favoring the study of more complex issues. Quantum mechanics stands
out as the theory of the $20^{th}$ century, shaping the most diverse
natural phenomena (from subatomic physics to cosmology). Yet, we are
lacking a foundational principle for quantum mechanics. In another
direction, the unification of quantum mechanics and general relativity,
string theory appears as the most promising example of a unified theory.
In a similar vain, we are lacking the conceptual foundation of string
theory.

In the present work we would like to further explore the suggestion
that the relational logic of C. S. Peirce may serve as the foundation
of both quantum mechanics and string theory {[}1{]}. It is most natural
to invoke logic in order to build and develop the theories describing
physical reality. The entire edifice of classical physics is grounded
on the Aristotelian logic, as it was transformed into algebraic logic
by G. Boole. The Boolean algebra is based on set theory. Set theory
is the backbone of the mathematical apparatus (continuum geometry,
differential equations) we employ in classical physics. Inadvertently
a set-theoretic enterprise is analytic, atomistic, arithmetic. Peirce
invented and elaborated a novel logical syntax, where relation is
the irreducible primary datum {[}2{]}. All other terms or objects
are defined in terms of relations, transformations, arrows, morphisms.
The Peircean logical structure bears great resemblance to category
theory {[}3{]}. A relational or categorical formulation is bound to
be synthetic, holistic, geometric. There is already an important work
in physics inspired by categorical notions {[}4,5,6{]}

The binary relation $R_{ij}$ between two {}``individual terms''
$S_{j}$ and $S_{i}$ may receive multiple interpretations: as a transition
from the $j$ state to the $i$ state, as a proof that the logical
proposition $j$ implies logical proposition $i$. At the very core
of the Peircean logical system is the composition of relations. Whenever
we have relations of the form $R_{ij}$, $R_{je}$ a third transitive
relation $R_{ie}$ emerges following the rule {[}2{]}
\begin{equation}
R_{ij}R_{ke}=\delta_{jk}R_{ie}\label{eq:1-1}
\end{equation}
The relations generate a $W_{\infty}$ algebra, which is a bosonic
extension of the Virasoro algebra, linked to area-preserving diffeomorphisms
{[}1{]}.

The simplified case of only two states $\left(i=1,2\right)$ indicates
the inherent nature of our logical system. Defining 
\begin{equation}
R_{z}=\frac{1}{2}\left(R_{11}-R_{22}\right)
\end{equation}
 and 
\begin{equation}
R_{+}=R_{12}\qquad R_{-}=R_{21}
\end{equation}
 we find out that the $SU\left(2\right)$ commutation relations are
satisfied {[}1{]}. Thus the underlying dynamics is similar to a {}``spin
$\nicefrac{1}{2}$ particle''. Considering that the answer to a logical
proposition is a {}``yes'' or a {}``no'' statement, analogous
to a \textquotedblleft{}spin up\textquotedblright{} or a \textquotedblleft{}spin
down\textquotedblright{} measurement, we view the spinor as the building
block of our logical construction. In this paper, following the Peircean
insight, we study relations built up by spinors and the deep connection
of spinors to geometry. We find out that composition of spinors gives
rise to Minkowski spacetime. On the other hand starting from geometry,
a string worldsheet or an $AdS_{3}$ spacetime, we recover the inherent
spinorial structure. At the end we comment on the hidden ties between
logic-quantum mechanics-string theory-geometry.

\subsubsection*{The emergence of Minkowski spacetime}

\qquad{}With $\left|r_{i}\right\rangle $ standing for a state or
a proposition, the suggested representation for $R_{ij}$ is {[}1{]}
\begin{equation}
R_{ij}=\left|r_{i}\right\rangle \left\langle r_{j}\right|
\end{equation}
where $\left\langle r_{i}\right|$ is the dual of $\left|r_{i}\right\rangle $.
Consider the spinor $\left|u\right\rangle =\left(\begin{array}{c}
\xi\\
\eta
\end{array}\right)$. The spinor relationship $R_{s}$ is defined by
\begin{equation}
R_{s}=\left|u\right\rangle \left\langle u\right|=\left(\begin{array}{cc}
\xi\xi^{*} & \xi\eta^{*}\\
\eta\xi^{*} & \eta\eta^{*}
\end{array}\right)
\end{equation}
The relationship $R_{s}$ receives the decomposition 
\begin{equation}
R_{s}=\frac{1}{2}\sum_{\mu=0}^{3}X_{\mu}\sigma^{\mu}\label{eq:1-6}
\end{equation}
with $\sigma_{0}=\mathbf{1}$ and $\sigma_{i}$ $\left(i=1,2,3\right)$
the Pauli matrices. We deduce that
\begin{eqnarray}
X_{0}= & \xi\xi^{*}+\eta\eta^{*} & =\left\langle u\right.\left|u\right\rangle \nonumber \\
X_{1}= & \xi\eta^{*}+\eta\xi^{*} & =\left\langle u\right|\sigma_{1}\left|u\right\rangle \\
X_{2}= & i\left(\xi\eta^{*}-\eta\xi^{*}\right) & =\left\langle u\right|\sigma_{2}\left|u\right\rangle \nonumber \\
X_{3}= & \xi\xi^{*}-\eta\eta^{*} & =\left\langle u\right|\sigma_{3}\left|u\right\rangle \nonumber 
\end{eqnarray}
$X_{\mu}$ satisfies identically the equation 
\begin{equation}
X_{0}^{2}=X_{1}^{2}+X_{2}^{2}+X_{3}^{2}\label{eq:1-8}
\end{equation}
and it may be viewed as a null vector belonging to Minkowski spacetime.
Thus the logic-algebraic origin of Minkowski spacetime becomes manifest.
It is Cartan who first indicated the profound links between spinors
and geometry {[}7{]} and Penrose who introduced spinors as a powerful
tool to address physics issues {[}8{]}.

For a normalized spinor with
\begin{equation}
\left|\xi\right|^{2}+\left|\eta\right|^{2}=1.0\label{eq:2-9}
\end{equation}
 the coordinates $X_{i}$ $\left(i=1,2,3\right)$ belong to the Bloch
sphere
\begin{equation}
X_{1}^{2}+X_{2}^{2}+X_{3}^{2}=1.0\label{eq:2-10}
\end{equation}
 We observe a mapping from the $S^{3}$ sphere, formed by the pair
of complex numbers satisfying eqn (\ref{eq:2-9}), to the $S^{2}$
sphere, eqn (\ref{eq:2-10}). It is the celebrated Hopf fibration
of $S^{3}$ by great circles $S^{1}$ and base space $S^{2}$ {[}9,10{]}.

A stereographic projection from the north pole to the plane $X_{3}=0$
will give the point $x_{1}+ix_{2}$ in the complex plane with
\begin{equation}
x_{1}+ix_{2}=\frac{X_{1}+iX_{2}}{1-X_{3}}=\frac{\xi^{*}}{\eta^{*}}\equiv\zeta^{*}
\end{equation}
 Another choice of the north pole will give another projection parametrized
by $\zeta'$, where 
\begin{equation}
\zeta'=\frac{a\zeta+b}{c\zeta+d}\label{eq:2-12}
\end{equation}
with $a,b,c,d$ complex numbers, subject to the condition $ad-bc\neq0$.
The M$\ddot{o}$bius transformation, eqn (\ref{eq:2-12}), is an automorphism
which preserves the conformal structure on the Bloch sphere. Specific
cases include dilation, rotation, translation, inversion.

Notice that the spinor $\left|u\right\rangle $ connected to the vector
$\vec{X}$ belonging to a Bloch sphere, equ. (\ref{eq:2-10}), creates
an oriented plane normal to the vector $\vec{X}$. The vectors $\vec{Y}\left(Y_{1},Y_{2},Y_{3}\right)$,
$\vec{Z}\left(Z_{1},Z_{2},Z_{3}\right)$ can be constructed as mutually
orthogonal, orthogonal to $\vec{X}$ and with unit norm. They are
defined by 
\begin{eqnarray}
Y_{1}+iZ_{1} & = & \xi^{2}-\eta^{2}\nonumber \\
Y_{2}+iZ_{2} & = & i\left(\xi^{2}+\eta^{2}\right)\label{eq:2-18}\\
Y_{3}+iZ_{3} & = & -2\xi\eta\nonumber 
\end{eqnarray}
 All spinor bilinears lie in this transverse plane.

Inversely we may attach to a null vector its spinor. Consider the
null momentum 
\begin{equation}
P_{0}^{2}=P_{1}^{2}+P_{2}^{2}+P_{3}^{2}
\end{equation}
 with the two solutions for $P_{0}$
\begin{eqnarray}
P_{0} & = & +\left(P_{1}^{2}+P_{2}^{2}+P_{3}^{2}\right)^{\frac{1}{2}}\label{eq:2-14}\\
P_{0} & = & -\left(P_{1}^{2}+P_{2}^{2}+P_{3}^{2}\right)^{\frac{1}{2}}\label{eq:2-15}
\end{eqnarray}
 The spinor equations corresponding to eqns (\ref{eq:2-14}), (\ref{eq:2-15})
are the Cartan-Weyl equations {[}11{]} 
\begin{eqnarray}
\left(\vec{P}\cdot\vec{\sigma}-P_{0}\right)\left|u_{+}\right\rangle  & = & 0\label{eq:2-16}\\
\left(\vec{P}\cdot\vec{\sigma}+P_{0}\right)\left|u_{-}\right\rangle  & = & 0\label{eq:2-17}
\end{eqnarray}
where now the spinors $\left|u\right\rangle $ are determined from
the components of the four-momentum $P_{\mu}$. The derivation suggests
that we may view a spinor as the square root of a null vector. The
Weyl spinors combine to provide the four-component Dirac spinor
\begin{equation}
\left|\psi_{D}\right\rangle =\left(\begin{array}{c}
u_{-}\\
u_{+}
\end{array}\right)
\end{equation}
 satisfying the massless Dirac equation {[}12{]}
\begin{equation}
\gamma^{\mu}\partial_{\mu}\left|\psi_{D}\right\rangle =0
\end{equation}
 
\begin{equation}
\gamma^{\mu}=\left(\begin{array}{cc}
0 & \sigma^{\mu}\\
\bar{\sigma}^{\mu} & 0
\end{array}\right)
\end{equation}
 with $\sigma^{\mu}=\left(1,\vec{\sigma}\right)$, $\bar{\sigma}^{\mu}=\left(1,-\vec{\sigma}\right)$.
We thus regard Dirac equation as the outcome of a logical construction.

\subsubsection*{Spinor reconstruction of the string worldsheet}

\qquad{}The defining operation of relation composition, eqn (\ref{eq:1-1}),
acquires another perspective when we represent the relation $R_{ij}$
by a double line {[}1{]}. Each state or proposition is represented
by a line, with a downward (upward) arrow attached to the initial
(final) state or proposition. The composition rule appears then as
string joining or string splitting. Repeated application of the composition
rule generates patterns derived within simplicial string theory and
reminding a triangulated Riemann surface {[}1{]}.

In the continuum the Polyakov string action is
\begin{equation}
S=-\frac{1}{2}T_{0}\int d\tau d\sigma\sqrt{-h}h^{\alpha\beta}\partial_{\alpha}X^{\mu}\partial_{\beta}X_{\mu}
\end{equation}
with $T_{0}$ the string tension, $h^{\alpha\beta}\left(\tau,\sigma\right)$
an auxiliary world-sheet metric and $X_{\mu}\left(\tau,\sigma\right)$
the parametrization of the string world-sheet {[}13{]}. Adopting a
flat world-sheet metric as a gauge choice, the solution of the string
equations is a sum of right-movers and left-movers
\begin{equation}
X^{\mu}\left(u,v\right)=X_{L}^{\mu}\left(u\right)+X_{R}^{\mu}\left(v\right)
\end{equation}
with $u=\tau+\sigma$, $v=\tau-\sigma$. The constraints become 
\begin{eqnarray}
\left(\partial_{u}X_{L}\right)^{2} & = & 0\label{eq:3-23}\\
\left(\partial_{v}X_{R}\right)^{2} & = & 0.\label{eq:3-24}
\end{eqnarray}
We consider string motion in three dimensions, the lowest dimension
for which string dynamics is not trivial. With $\mu=0,1,2$ and defining
$P_{Li}=\partial_{u}X_{L}^{i}$ the constraint (\ref{eq:3-23}) becomes
\begin{equation}
P_{L0}^{2}=P_{L1}^{2}+P_{L2}^{2}.\label{eq:3-25}
\end{equation}
This is the celebrated Pythagoras equation. Pythagoras' equation is
derived from the geometrical requirement of composing two squares
into a third square. We may attempt a linear representation of the
quadratic Pythagorean form by writing
\begin{equation}
R_{P}=P_{L0}\sigma_{0}+P_{L1}\sigma_{1}+P_{L2}\sigma_{3}=\left(\begin{array}{cc}
P_{L0}+P_{L2} & P_{L1}\\
P_{L1} & P_{L0}-P_{L2}
\end{array}\right)
\end{equation}
 Equation (\ref{eq:3-25}) becomes then $det\, R_{P}=0$. $R_{P}$
viewed as a relationship of a real spinor $\left|\phi\right\rangle =\left(\begin{array}{c}
p\\
q
\end{array}\right)$ takes the form
\begin{equation}
R_{P}=\left|\phi\right\rangle \left\langle \phi\right|=\left(\begin{array}{cc}
p^{2} & pq\\
pq & q^{2}
\end{array}\right)
\end{equation}
 We deduce then
\begin{eqnarray}
P_{L0} & = & \frac{1}{2}\left(p^{2}+q^{2}\right)\\
P_{L1} & = & pq\\
P_{L2} & = & \frac{1}{2}\left(p^{2}-q^{2}\right)
\end{eqnarray}
The above parametrization is a zero Hopf map with $p$ and $q$ the
coordinates of $\mathbb{R}^{2}$. We may notice also that through
a physical spinorial process we managed to enact the geometrical theorem
of Pythagoras. In a similar fashion we represent the right-movers
by two real functions $r$ and $s$. Thus the entire string world-sheet
is described by four independent dynamical variables $\left(p,q,r,s\right)$.
Within our approach the parametrization of the string world-sheet
becomes
\begin{eqnarray}
X^{0}\left(u,v\right) & = & \frac{1}{2}\left[\intop^{u}\left(p^{2}+q^{2}\right)du+\intop^{v}\left(r^{2}+s^{2}\right)dv\right]\\
X^{1}\left(u,v\right) & = & \intop^{u}pq\, du+\intop^{v}rs\, dv\\
X^{2}\left(u,v\right) & = & \frac{1}{2}\left[\intop^{u}\left(p^{2}-q^{2}\right)du+\intop^{v}\left(r^{2}-s^{2}\right)dv\right]
\end{eqnarray}
 We recognise that the above parametrization is the Enneper - Weierstrass
parametrization of a minimal surface {[}14,15{]}.

String theory appears as the most promising framework for unifying
quantum mechanics and gravity, for creating quantum gravity. Our analysis
points out the affinity between quantum mechanics and string theory,
since both reside on the same relational principles. The known similarity
of string theory to two dimensional gravity, further strengthens our
confidence to string theory as a candidate theory for quantum gravity.
As an extra dividend, our spinor reconstruction of the string dynamics
solves effectively the constraints and allows us to use only the real,
independent degrees of freedom.

\subsubsection*{Building AdS}

\qquad{}Anti-de Sitter spacetime attracted enormous attention as
the locus of a duality between gravity living in the interior of AdS
and a conformal field theory living in the boundary of AdS {[}16,17,18{]}.
It is quite natural to study the internal symmetries and architecture
of AdS, in order to better discern the inherent dynamics. AdS space
is the maximally symmetric solution of Einstein's equations with an
attractive cosmological constant. Anti-de Sitter spacetime is defined
as a quadric surface embedded in a flat space of signature $\left(++-\cdots-\right)$.
Thus $AdS_{3}$ is defined as the hypersurface
\begin{equation}
u^{2}+v^{2}-x_{1}^{2}-x_{2}^{2}=1.0
\end{equation}
embedded in a four-dimensional flat space with the metric 
\begin{equation}
ds^{2}=du^{2}+dv^{2}-dx_{1}^{2}-dx_{2}^{2}.
\end{equation}
We may proceed to a spinorial reconstruction of $AdS_{3}$. Starting
with a spinor $\left|\varphi\right\rangle =\left(\begin{array}{c}
\varphi_{1}\\
\varphi_{2}
\end{array}\right)$, the {}``dual'' spinor $\left|\tilde{\varphi}\right\rangle =\eta\left|\varphi\right\rangle $
where $\eta=\left(\begin{array}{cc}
1 & 0\\
0 & -1
\end{array}\right)$, the inner product becomes 
\begin{equation}
\left\langle \tilde{\varphi}|\varphi\right\rangle =\left|\varphi_{1}\right|^{2}-\left|\varphi_{2}\right|^{2}.\label{eq:23}
\end{equation}
Defining $\varphi_{1}=a+ib$, $\varphi_{2}=c+id$, the normalization
condition $\left\langle \tilde{\varphi}|\varphi\right\rangle =1.0$
provides
\begin{equation}
a^{2}+b^{2}-c^{2}-d^{2}=1.0
\end{equation}
The obvious identification $a=u$, $b=v$, $c=x_{1}$, $d=x_{2}$
reproduces $AdS_{3}$.

In analogy with the Hopf maps of a sphere to a sphere, we may look
for the non-compact Hopf map of a hyperboloid. Let us define
\begin{equation}
\Sigma_{1}=i\sigma_{1}\qquad\Sigma_{2}=i\sigma_{2}\qquad\Sigma_{3}=\sigma_{3}
\end{equation}
We obtain then
\begin{eqnarray}
y_{1}=\left\langle \tilde{\varphi}\left|\Sigma_{1}\right|\varphi\right\rangle  & = & i\left(\varphi_{1}^{*}\varphi_{2}-\varphi_{2}^{*}\varphi_{1}\right)\nonumber \\
y_{2}=\left\langle \tilde{\varphi}\left|\Sigma_{2}\right|\varphi\right\rangle  & = & \varphi_{1}^{*}\varphi_{2}+\varphi_{2}^{*}\varphi_{1}\\
y_{3}=\left\langle \tilde{\varphi}\left|\Sigma_{3}\right|\varphi\right\rangle  & = & \varphi_{1}^{*}\varphi_{1}+\varphi_{2}^{*}\varphi_{2}\nonumber 
\end{eqnarray}
We observe that 
\begin{equation}
y_{3}^{2}-y_{1}^{2}-y_{2}^{2}=\left[\left|\varphi_{1}\right|^{2}-\left|\varphi_{2}\right|^{2}\right]^{2}=1.0
\end{equation}
The above equation describes the hyperboloid $H^{2,0}$ and if we
accept imaginary coordinates it is equivalent to $AdS_{2}$. Thus
we observe the Hopf fibration $AdS_{3}\rightarrow AdS_{2}$ and $AdS_{3}$
may be written in the form of a bundle over $AdS_{2}$ {[}19,20{]}.

We may seek a transformation of the spinor $\left|\varphi\right\rangle $,
$\left|\varphi'\right\rangle =R\left|\varphi\right\rangle $, leaving
the form of eqn (\ref{eq:23}) invariant
\begin{equation}
\left\langle \tilde{\varphi}'|\varphi'\right\rangle =\left\langle \tilde{\varphi}|\varphi\right\rangle .
\end{equation}
The transformation $R$ takes the form 
\begin{equation}
R=\left(\begin{array}{cc}
\cosh\omega\, e^{i\theta} & \sinh\omega\, e^{-i\chi}\\
\sinh\omega\, e^{i\chi} & \cosh\omega\, e^{-i\theta}
\end{array}\right).
\end{equation}
Two succesive $R$ transformations generate a third transformation.
Let us represent the transformation by $r\left(\alpha,\beta\right)$,
where $\alpha=\cosh\omega\, e^{i\theta}$ and $\beta=\sinh\omega\, e^{-i\chi}$.
Then with $r_{1}\left(\alpha,\beta\right)$ and $r_{2}\left(\gamma,\delta\right)$
the succesive transformations, we find
\begin{equation}
r_{1}r_{2}=r_{3}\left(\alpha\gamma+\beta\delta^{*},\alpha\delta+\beta\gamma^{*}\right).
\end{equation}
The inverse of a transformation $r\left(\alpha,\beta\right)$ is the
transformation $r^{-1}\left(\alpha^{*},-\beta\right)$.

Writing $\zeta=\frac{\varphi_{1}}{\varphi_{2}}$ the transformation
becomes
\begin{equation}
\zeta'=\frac{\alpha\zeta+\beta}{\beta^{*}\zeta+\alpha^{*}}.\label{eq:31}
\end{equation}
Equation (\ref{eq:31}) is the M$\ddot{o}$bius transformation which
preserves the conformal structure of $AdS_{3}$

\subsubsection*{In lieu of conclusions}

\qquad{}We suggested relational logic as the appropriate framework
to comprehend relational systems, notably quantum mechanics and string
theory. A logical proposition receiving a {}``yes'' or a {}``no''
answer, may be represented by a spinor. A composition of logic spinors,
ressembling to a spinor network, would be equivalent to a logical
proof. In the present work we considered the simplest spinor compositions
and how they lead to the representation of Minkowski spacetime, $AdS_{3}$
spacetime and the string world-sheet. Geometry and spacetime, rather
than abstract methematical constructions, emerge as the outcome of
a logical process. Our demarche is reminiscent of the Wheeler's pioneering
idea of pregeometry {[}21,22{]}. Wheeler pointed out that the genuine
explanations about the nature of something do not come about by explicating
a concept in terms of similar ones, but by reducing it to a different,
more basic kind of object. Citing Peirce, Wheeler considers that law
must have come into being and that physics must be built from a foundation
that has no physics {[}23{]}. Taking account of the most challenging
{}``experimental'' fact, that the universe is comprehensible, we
should build the universe on this very demand for comprehensibility.
Within this perspective, pregeometry (the stage preceding geometry)
is based on the calculus of propositions {[}23{]}. We consider our
approach as a concrete realization of Wheeler's vision, bringing together
logic, geometry, string theory, quantum mechanics. The common thread
of these theories is their shared inner syntax, the relational logic
of Peirce.

The holographic principle, developed by 't Hooft {[}24{]} and Susskind
{[}25{]} after an analysis of the black hole dynamics, dictates that
the degrees of freedom contained in a volume $V$ are encoded on the
surface bounding $V$ like a holographic image. A detailed study case
is given in {[}26{]}. An oscillating string, encircling a black hole
is contracting towards the horizon. String fluctuations are developed
in both the radial and angular directions. To a static asympotic observer
the radial fluctuations are suppresed by the Lorentz-contraction and
thus the string appears eventually to cover the whole black hole horizon
{[}26{]}. The identification of the horizon with the string world-sheet
gives credit to the notion that the stringy degrees of freedom are
registered on the entire area of the event horizon. The two-dimensional
horizon may be seen as a spherical holographic screen and an infinitesimal
pixel on such a screen may be represented by a null vector $X_{\mu}$,
equ. (\ref{eq:1-8}) with $X_{0}=1$. The same spinor giving giving
rise to $\vec{X}$, creates also the plane transverse to $\vec{X}$,
equ. (\ref{eq:2-18}). The spinor holographic construction can be
extended to $d$ dimensions {[}27{]} and it appears that pure spinors
are best suited for an analysis of the AdS/CFT holography {[}28{]}.

We studied relations built up by spinors and leading to geometrical
patterns. We considered relations generated by a single spinor. It
is very interesting to consider relations built by two spinors, since
such a construction will lead to a better comprehension of the quantum
entanglement. We may even go further and consider, following Wheeler's
advice, the statistical analysis of a great number of logical-spinor-propositions.
If this approach makes sense, the entire universe will look like a
theorem. Work along these lines is in progress.

\part*{Acknowledgments}

This work was supported by the Templeton Foundation through the project
{}``Quanto-Metric''.

\end{document}